 \documentclass[final,3p,twocolumn]{elsarticle}

 \usepackage{graphicx}

\usepackage{amssymb}

\journal{Nuclear Instruments and Methods in Physics Research Section A}

\begin{document}

\begin{frontmatter}



\title{The time-of-propagation counter for Belle II}


\author{K.~Nishimura on behalf of the Belle II Particle Identification Group}

\address{Department of Physics and Astronomy, University of Hawaii, 2505 Correa Road, Honolulu, HI 96822, USA}

\begin{abstract}
The Belle II detector operating at the future upgrade to the KEKB accelerator 
will perform high-statistics precision investigations into the flavor sector of
the Standard Model. As charged hadron identification is a vital element of the 
experiment's success, the time-of-propagation (TOP) counter has been chosen as 
the primary particle identification device in the barrel region of Belle II. 
The TOP counter is a compact variant of the detection of internally reflected 
Cherenkov light (DIRC) technique and relies heavily on exquisite single photon 
timing resolution with micro-channel plate photomultiplier tubes. We discuss 
the general principles of TOP operation and optimization of the Belle II TOP 
configuration, which is expected to provide 4 sigma or better separation of 
kaons and pions up to momenta of approximately 4 GeV$/c$.
\end{abstract}


\end{frontmatter}


\section{Introduction}
\label{sec:Introduction}

The Belle \cite{Belle} and BaBar \cite{BaBar} experiments at the 
KEKB and PEP-II $B$ factories have collected a combined integrated luminosity 
of over 1.5 ab$^{-1}$.  This wealth of data has resulted in a number of 
precision measurements, including confirmation of the Kobayashi-Maskawa 
mechanism of $CP$ violation. Charged 
particle identification (PID) systems have played a significant role in 
both experiments by providing discrimination between $K^\pm$ and $\pi^\pm$, and 
hence enhancing efficiencies for detecting rare $B$ decays as well as 
improving the flavor tagging necessary for extracting time-dependent $CP$ 
asymmetries. In order to further probe the Standard Model and search for 
new physics, detectors at the next generation super $B$ factories will need 
increasingly precise PID devices that can accomodate larger background 
rates due to significantly higher luminosities.

To deal with these challenges, the Belle II detector will replace the 
existing barrel region time-of-flight counter and threshold aerogel Cherenkov 
counter with a time-of-propagation (TOP) counter.  The TOP technique is a 
compact variant of the detection of internally reflected Cherenkov light 
(DIRC) method \cite{BaBar_DIRC}, wherein Cherenkov photons emitted as a 
charged particle passes through a radiator material, such as a fused-silica 
bar, are carried by total internal reflection to the bar's end and detected on 
an imaging plane.  In contrast to the BaBar DIRC, which 
uses two dimensional imaging on a $\sim$m$^2$ image plane along with $\sim$ns 
timing to reject backgrounds, the TOP counter will use a compact 
$\sim$100 cm$^2$ image plane near the bar end, and further use micro-channel 
plate photomultiplier tubes (MCP-PMTs) with timing precision of $\sim$40 ps to 
further distinguish particle types.  We describe the proposed geometry of the 
Belle II TOP detector, including the choice of MCP-PMTs and readout 
electronics, as well as simulated performance for $K/\pi$ separation.

\section{Principle of Operation}
\begin{figure}
\includegraphics[width=7.8cm]{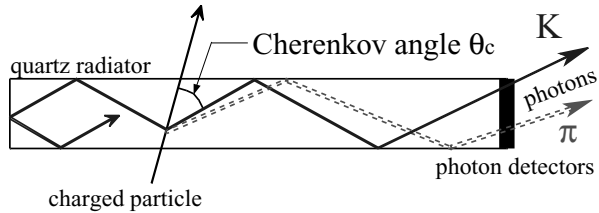}
\caption{A schematic side-view of a TOP counter.}
\label{fig:concept}
\end{figure}

The basic TOP counter schematic is shown in Fig. \ref{fig:concept}
\cite{TOP_RICH2007}. A charged 
particle of a given momentum, which is measured by the tracking subdetectors, 
passes through a radiator material, emitting Cherenkov photons in a cone with 
opening angle, $\theta_C$, determined by the particle velocity, $\beta$.  Thus, 
the characteristic cone will be unique for particles of different mass.  
A fraction of the photons are transported down the bar via total internal 
reflection, and the characteristic pattern is measured in time and one or more 
spatial dimensions by PMTs mounted directly to the bar end.  The 
time-of-propagation of each photon from its emission point to the detector 
plane is correlated with $\theta_C$.  Additionally, photon arrival times are 
typically measured relative to the time of the collider event, so 
photons are further separated in time by the time-of-flight of the species of 
charged particle.  

\begin{figure}[htb]
\includegraphics[width=7.8cm]{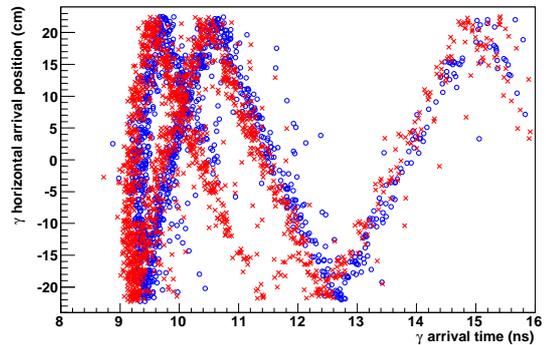}
\caption{A typical hit pattern in the horizontal bar dimension versus arrival 
time for a simple TOP counter, consisting of an ensemble of 100 events each for 
incident pions (red crosses) and kaons (blue circles).}
\label{fig:concept-simulation}
\end{figure}

The simplest TOP counter reads out only information in one
spatial direction and in time. A typical hit pattern for this type of 
detector is shown in Fig. \ref{fig:concept-simulation}. Arrival time 
differences between photons from kaons and pions for any given channel differ 
by order $\sim$100 ps at 2 GeV$/c$, and less at higher momenta.  For best 
performance, the timing resolution of any TOP photodetector must be 
significantly smaller than this separation. 
The overall time resolution is limited by 
chromatic dispersion in the radiator material.  To mitigate this effect, a 
filter material can be placed in front of the MCP-PMTs, limiting 
detected wavelengths to those where chromatic effects are less pronounced. 
A focusing mirror can also be added to one end of the radiator bar 
\cite{focusing}.  Along 
with finer segmentation in the other spatial direction, this allows some 
photons of different colors, and thus slightly different $\theta_C$, to focus 
on different photodetector channels.  The focusing element can also 
remove ambiguities in the photon emission point due to the finite bar 
thickness.  At the cost of some compactness, the image plane can be expanded 
beyond the end of the bar for improved spatial resolution \cite{TIPP09_iTOP}.

\begin{figure}[htb]
\centerline{\includegraphics[width=8cm]{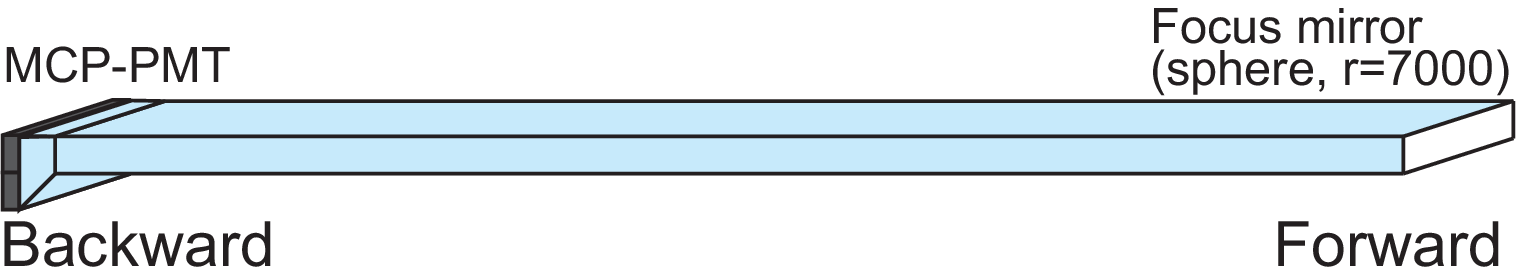}}
\centerline{\includegraphics[width=8cm]{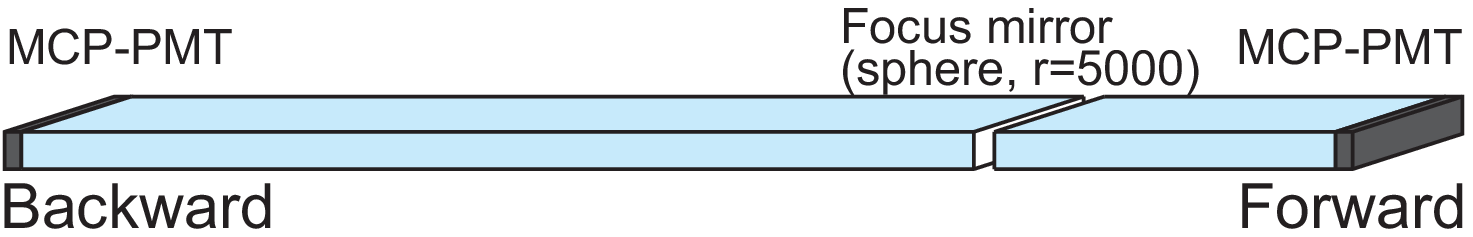}}
\caption{Conceptual sketches of the 1-bar and 2-bar TOP baseline designs.}
\label{fig:baselinedesign}
\end{figure}

\section{Detector Configuration}
\label{sec:Design}
Two designs have been studied for use in the Belle II detector, each shown 
schematically in Fig. \ref{fig:baselinedesign}.  In both cases the design 
consists of a barrel of 16 TOP modules.  The radiator cross section is 
approximately 440 mm wide by 20 mm thick.  Because there is 
no strong optimum in performance for the width of the bar, the chosen width 
represents a reasonable compromise between manufacturing cost, mechanical 
stability, and overall acceptance. The thickness represents a trade-off 
between increased light yield and a larger mass in front of the calorimeter.

The 1-bar design, also referred to as an imaging TOP (iTOP), 
employs a radiator of length $\sim$ 2.75 m, with a spherical 
mirror placed on the forward end and an expansion prism on the backward end, 
resulting in an image plane approximately twice the size of the bar 
cross-section.  This creates a larger image plane and thus clarifies the 
ring image and improves the wavelength separation.  In the 2-bar design, two 
radiators are used.  The ``long'' radiator in the backward direction is 
reduced in length relative to the 1-bar design, reducing the effect of 
chromatic dispersion.  As in the 1-bar design, this radiator utilizes a 
spherical mirror in the forward direction.  The two bars are read out directly 
at the ends, resulting in two separate detector planes, each of size 
equivalent to the bar cross section.  In both designs, a wavelength filter of 
$\lambda \gtrsim $ 400 nm is introduced in front of the PMTs to further reduce 
chromatic effects.

The full detector consists of a barrel of 16 modules. Due to gaps between the 
sides of each bar to accommodate the expansion volume and/or support 
structures, the azimuthal acceptance for charged tracks is $\sim$95\%. 

\section{Photodetectors and Readout Electronics}
\begin{figure}[htb]
\centerline{
 \includegraphics[height=3.5cm]{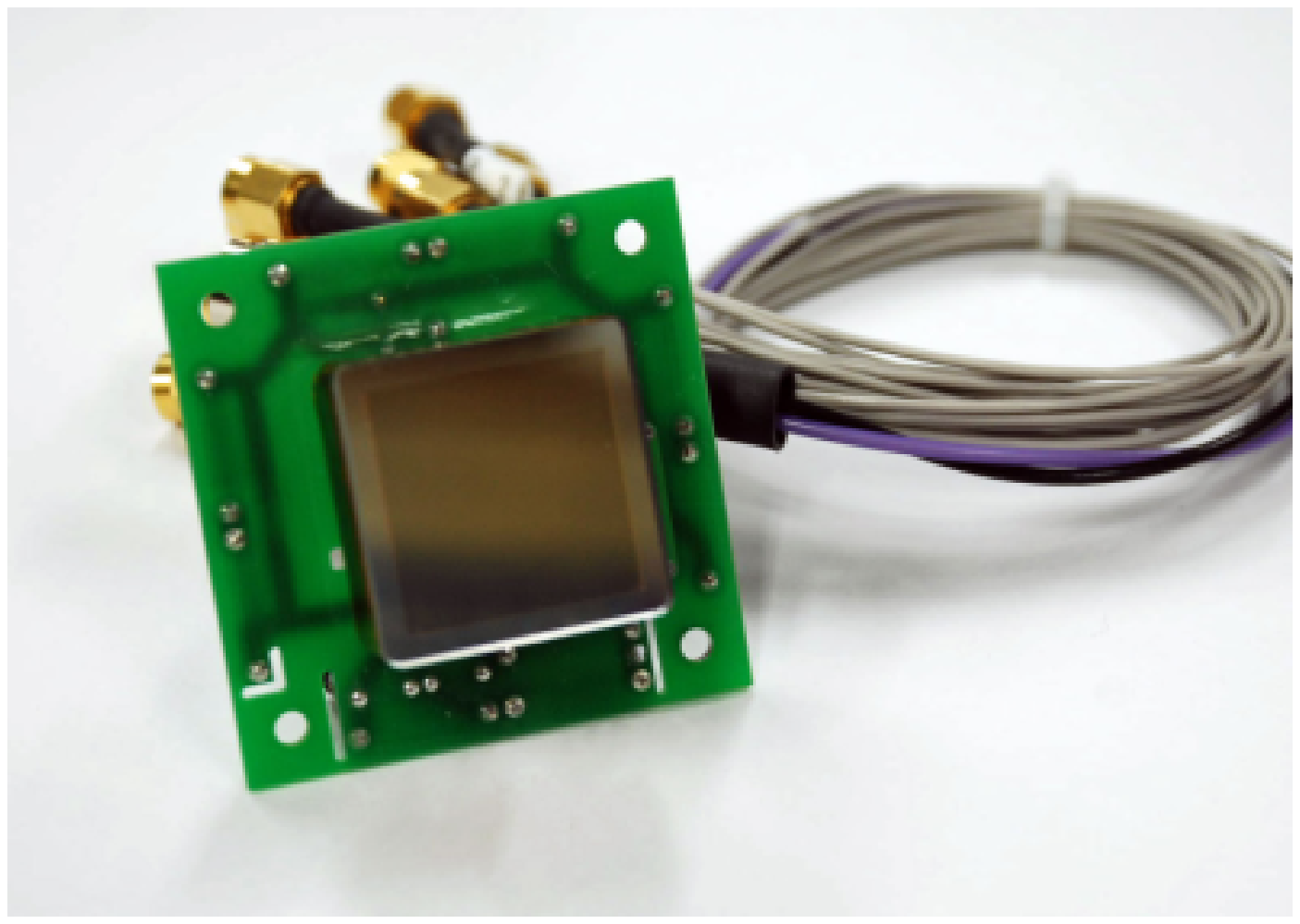}
 \includegraphics[height=3.5cm]{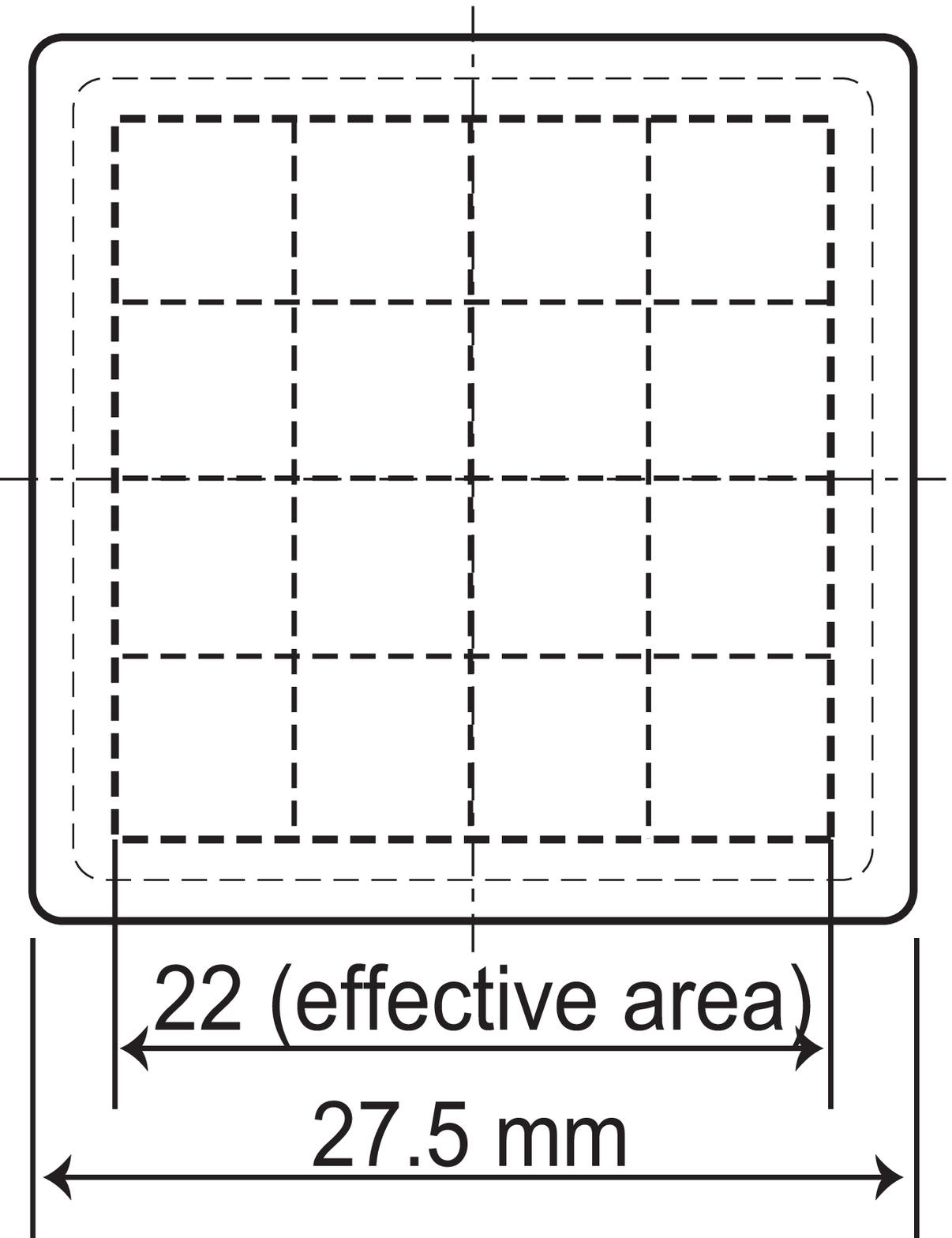}}
 \caption{SL-10 prototype MCP-PMT photograph (left) and anode
 structure diagram (right).}
\label{fig:SL-10}
\end{figure}

The baseline MCP-PMT for the Belle II TOP is the Hamamatsu SL-10 
\cite{SL10}, shown in Fig. \ref{fig:SL-10}.  The SL-10 has a pore size of 
10 $\mu$m and a single photon timing resolution of $\sim$40 ps.  An aluminum 
protection layer on the second of two MCPs protects against ion feedback and 
prevents premature aging of the photocathode. A 4x4 anode 
configuration will be used, resulting in a total of 512 channels per TOP 
module, or $\sim$8200 channels for the entire barrel.  Existing 
prototypes utilize a multi-alkali photocathode, but efforts are currently 
underway to switch to a super bi-alkali photocathode and increase overall 
quantum efficiency.

The large number of readout channels and space constraints of the detector 
necessitate compact front-end electronics. To accomodate this need while 
maintaining high precision timing, the Belle II TOP counter will utilize a 
waveform sampling application specific integrated circuit (ASIC) for readout 
of the MCP-PMTs.  Timing precision of $\leq$ 30 ps with an analog bandwidth of 
approximately 1 GHz has been obtained with the existing LABRADOR3 ASIC 
\cite{LAB3}.  Belle II will utilize a Buffered LABRADOR (BLAB) ASIC, 
which adds front-end amplification for detection of single 
photo-electron signals and significantly deeper sampling to accomodate 
the expected trigger latencies of up to 5 $\mu$s.

Two generations of BLAB ASICs have been tested with the fDIRC prototype.  The
first generation BLAB1 \cite{BLAB1} was utilized for 16 channels of PMT 
readout in a beam test of the focusing DIRC (fDIRC) 
prototype \cite{fDIRC_Beam_Test}, where it obtained timing resolutions 
competitive with conventional constant fraction discrimination techniques.  
The second generation BLAB2 has been used to instrument the same fDIRC 
prototype at a cosmic ray test stand \cite{fDIRC_CRT}.  In this case a 
compact, integrated photodetector readout was developed and implemented, 
allowing readout of approximately 450 channels. The TOP detector will utilize 
the BLAB3 ASIC, with specifications listed in Table \ref{tab:BLAB3_specs}.  

\begin{table}[h]
\caption{Required specifications for the BLAB3 readout ASIC.}
\label{tab:BLAB3_specs}
\begin{center}
\resizebox{!}{2.25 cm}{
\begin{tabular}{ll}
Parameter & Value \\ \hline \hline
Channels/BLAB3 & 8 \\
Analog bandwidth & $\geq$ 500 MHz\\
Sampling speed & 4 Giga-samples/second\\
Samples/channel & 32768\\
Amplifier gain & 60 relative to 50 $\Omega$ signal\\
Trigger channels & 8\\
Effective resolution & $\approx $9 bits\\
Sample convert window & 64 samples ($\approx $16~ns) \\
Readout granularity & 1 sample, random access \\
Readout time for $n$ samples & $1+n \cdot 0.02$ $\mu$s\\
Sustained level 1 trigger rate & 30 kHz\\ \hline \hline
\end{tabular}}
\end{center}
\end{table}

Information from the front-end will be transferred to the rest of the Belle II 
data acquisition and trigger systems via fiberoptic links.  Back-end  
electronics provide online waveform processing, performing feature extraction 
on full waveforms and reducing the large raw data volume to a manageable size.

\section{Expected Performance}

Monte Carlo simulations have been performed to estimate the $K/\pi$ separation 
performance of the proposed TOP geometries.  Three independent simulations 
have been performed.  The first utilizes GEANT3 \cite{GEANT3} to generate the 
initial charged tracks as well as simulate electromagnetic and hadronic 
interactions; a stand-alone routine is then used to track 
optical photons to their final detection at the image plane.  The second uses 
Geant4 \cite{Geant4} to initiate the primary track, simulate electromagnetic 
interactions, and perform optical propagation to the detector plane.  The last 
utilizes stand-alone code to simulate Cherenkov emission in the bar and track 
photons to the detector plane; in this set of simulations no secondary 
interactions are included. All three simulations include measured 
distributions for the time and spectral response of the MCP-PMTs.  Beam 
tests have been performed \cite{TOP_Beam_Test} and validate the simulation 
results.

To determine expected efficiencies and fake rates for $K/\pi$ discrimination, 
the detected photons for each track are tested against probability 
distribution functions (PDFs) for each particle hypothesis 
($\mathcal{P}^K(x,t)$ and $\mathcal{P}^\pi(x,t)$).  From these PDFs, a 
likelihood is determined for a simulated primary charged particle at a given 
incident angle and impact position.
For the Geant
based simulations, PDF values are obtained from an ensemble of a large number 
of simulated events.  For the fully stand-alone code, an analytical technique 
is used to calculate PDF values \cite{TOP_Reconstruction}.

\begin{figure}[htb]
\centerline{\includegraphics[width=8cm]{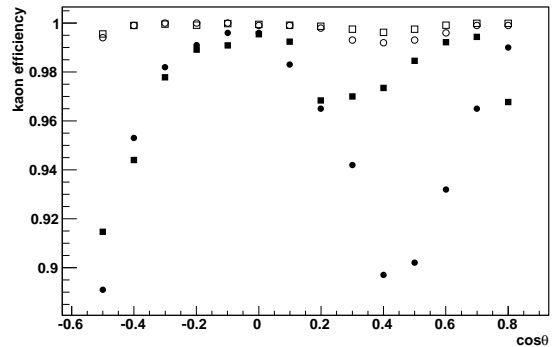}}
 \caption{Geant4-based simulated kaon efficiency as a function of cosine of 
track polar angle for the 1-bar (circles) and 2-bar (squares) configurations 
at track momenta of 2 GeV/$c$ (open) and 3 GeV/$c$ (filled). }
 \label{fig:Geant4_eff_fake}
\end{figure}

Example distributions of kaon efficiency for the Geant4 
simulations are shown in Fig. \ref{fig:Geant4_eff_fake}.  The 
three simulations show reasonable agreement, particularly in terms of 
reproducing similar features as a function of track impact angle.  A dip in 
performance is observed for the 1-bar design at a polar angle 
of $\sim$60$^\circ$, and is attributed to chromatic dispersion due to the 
increased propagation length of the photons.  The same feature is seen in the 
2-bar design around $\sim$70$^\circ$, but is less pronounced due to the shorter 
bar length.  The two bar design also shows a noticeable drop in performance 
in the very forward direction.  In this region, the short forward bar serves 
primarily as a time-of-flight counter, since the characteristic ring images 
do not separate significantly temporally or spatially before detection.  
Despite these differences between configurations, the expected overall 
differences for representative physics modes are mild.  For example, 
Table \ref{tab:eff_fake_rhogamma} shows peformance for the decay 
$B \to \rho \gamma, \rho \to \pi\pi$, which suffers from a large background 
from $B \to K^* \gamma, K^* \to K \pi$.  Differences in pion efficiencies 
and kaon misidentification rates for the two TOP designs are less than a 
percent in all simulations.  Efforts are continuing to improve simulations, 
add all appropriate effects, and test the analytical reconstruction method 
on GEANT3 and Geant4 simulation data.

\begin{table}[htb]
  \begin{center}
  \caption{
  Simulated pion efficiencies, $\epsilon_\pi$, and kaon misidentification rates,
 $f_K$, for the decay $B \to \rho\gamma$ with a background of 
  $B \to K^* \gamma$, under nominal expected conditions. }
      \label{tab:eff_fake_rhogamma}
    \begin{tabular}{lcccc}
      \hline
      \hline
   & \multicolumn{2}{c}{$\epsilon_\pi$ (\%)} & \multicolumn{2}{c}{$f_K$ (\%)}\\
      \hline
                  & 1-bar & 2-bar & 1-bar & 2-bar \\
      GEANT3      & 98.6  & 98.9  &  0.9  &  1.0  \\
      Geant4      & 99.9  & 99.9  &  0.4  &  0.2  \\
      Stand-alone & 99.1  & 99.6  &  0.4  &  0.4  \\
      \hline
      \hline
    \end{tabular}
  \end{center}
\end{table}

These simulations have been used to study various effects and uncertainties. 
Small or negligible effects are observed for possible mirror misalignment and 
beam background rates.  More significant effects are observed due to 
uncertainties from the tracking systems and jitter in the measurement of the 
event start time, $t_0$. The 1-bar design is more sensitive to the former and 
the 2-bar design to the latter, though the two designs otherwise perform 
comparably. While the expected tracking uncertainties are relatively well 
understood, the final expected $t_0$ jitter is much less clear.  For this 
reason, as well as because of the larger acceptance in the forward direction 
and the simplicity of a single detector plane, the 1-bar iTOP has been 
chosen as the baseline design.

\section{Conclusion}

Belle II will utilize a TOP counter for PID in the barrel region, as its 
compact design means it can be integrated into the limited space available in 
the Belle II environment. Two designs have been studied: a 2-bar configuration 
with two readout planes and a 1-bar configuration with a single, larger image 
plane.  Both configurations utilize the Hamamatsu SL-10 MCP-PMT as the 
photodetector, as this device meets the extreme timing requirements required 
for the TOP technique. Electronics based on a custom waveform sampling ASIC, 
the BLAB3, will perform front-end readout of the photodetectors and interface 
to the back-end data and trigger system.  Simulation of both configurations 
indicates tradeoffs between the two geometries for various event uncertainties 
and impact positions. In 
both cases, performance is expected to surpass the existing Belle PID systems, 
allowing 4 sigma separation of kaons and pions up to momenta of 4 GeV$/c$.





\bibliographystyle{elsarticle-num}

\begin{thebibliography}{00}


\bibitem{Belle}
A.~Abashian {\it et al.} (Belle Collaboration),
Nucl. Instrum. Meth. A {\bf 479}, 117 (2002).

\bibitem{BaBar}
B.~Aubert {\it et al.} (BaBar Collaboration),
Nucl. Instrum. Meth. A {\bf 479}, 1 (2002).

\bibitem{BaBar_DIRC}
I.~Adam, {\it et al.} (BaBar-DIRC Collaboration), 
Nucl. Instrum. Meth. A {\bf 538}, 281 (2005).

\bibitem{TOP_RICH2007}
K.~Inami, 
Nucl. Instrum. Meth. A {\bf 595}, 96 (2008).

\bibitem{focusing}
B.~Ratcliff,
Nucl. Instrum. Meth. A {\bf 502}, 211 (2003).

\bibitem{TIPP09_iTOP}
K.~Nishimura, {\it et al.}, 
Nucl. Instrum. Meth. A, DOI: 10.1016/j.nima.2010.02.227.

\bibitem{SL10}
K.~Inami, {\it et al.}, 
Nucl. Instrum. Meth. A {\bf 592}, 247 (2008).

\bibitem{LAB3}
G.~Varner, {\it et al.},
Nucl. Instrum. Meth. A {\bf 583}, 447 (2007).

\bibitem{BLAB1}
L.~Ruckman {\it et al.}, 
Nucl. Instrum. Meth. A {\bf 591}, 534 (2008).

\bibitem{fDIRC_Beam_Test}
J.~Benitez {\it et al.},
Nucl. Instrum. Meth. A {\bf 595}, 104 (2008).

\bibitem{fDIRC_CRT}
L.~Ruckman {\it et al.},
Nucl. Instrum. Meth. A, DOI: 10.1016/j.nima.2010.02.229.

\bibitem{GEANT3}
R. Brun {\it et al.}, GEANT 3.21, CERN Report Number DD/EE/84-1 (1984).

\bibitem{Geant4}
S.~Agostinelli {\it et al.}, 
Nucl. Instrum. Meth. A {\bf 506}, 250 (2003).

\bibitem{TOP_Beam_Test}
K.~Inami,
Journal Intr. {\bf 5} P03006 (2010).

\bibitem{TOP_Reconstruction}
M.~Staric {\it et al.},
Nucl. Instrum. Meth. A {\bf 595}, 252 (2008).

 \end{thebibliography}



\end{document}